\input epsf
\documentstyle[preprint,prb,aps]{revtex}

\begin{document}
\draft
\preprint{cond-mat/9410021 revised 10/17}

\title{
Model for FeSi,
a strongly correlated insulator
}

\author{Castor Fu}
\address{
Dept. of Physics, Stanford University, Stanford, CA 94305\\
and Theoretical Division, MS B213, Los Alamos National  Laboratory,
Los Alamos, NM 87544
}
\author{S. Doniach\cite{byline}}
\address{
Dept. of Applied Physics, Stanford University, Stanford, CA 94305
}
\date{October 8, 1994}
\maketitle

\begin{abstract}
Prompted by suggestions that FeSi  may be a strongly correlated insulator,
somewhat analogous to the Kondo insulators, we study the properties of a
two band Hubbard model Hamiltonian in the infinite dimensional limit.
By a calculation of the one-particle self energy using
self consistent
second order perturbation theory
we are able to calculate the temperature dependence of the magnetic
susceptibility and of angle resolved photoemission spectra.  The
results are in fair agreement with experimental data on FeSi.

\end{abstract}

\narrowtext
\section{Introduction}
\label{sec:intro}
With the renewed interest in strongly correlated insulators, FeSi has been
the subject of  renewed experimental and theoretical examination.
\cite{Schlesinger93,Mattheiss93,Fu94,Sarrao94,Continentino94}
Infrared reflectivity \cite{Schlesinger93}
and angle resolved photoemission \cite{Shen94} measurements are providing
a more detailed
view of the effects of strong correlations on the electronic
excitations in FeSi.  In light of these experiments, we have
constructed a simplified model for a strongly correlated insulator
where the correlation effects can be easily calculated.

In early research, FeSi was identified as a narrow ($\approx 50$ meV) gap
semiconductor with unusual magnetic properties.
Its magnetic susceptibility was recognized as
inexplicable from simple band structure models,\cite{Jaccarino67}
which would require extremely narrow bands.  However, there were no signs
of the magnetic ordering which one might expect from such localized
states
at low temperatures.
Since then, explanations for the magnetic behavior in terms of
spin fluctuations\cite{Moriya83,Evangelou83} have had reasonable success
and have even predicted the existence of a temperature induced magnetic
moment, which was found experimentally only much later. \cite{Tajima88}

The discovery of the high temperature superconductors stimulated new
interest in the subject of strongly correlated insulators such as
CeNiSn, Ce$_3$Bi$_4$Pt$_3$ and SmB$_6$ and FeSi.  In the case of FeSi,
electronic structure calculations based on the local density
approximation (LDA)\cite{Mattheiss93,Fu94} find an insulating ground
state and  elastic properties in reasonable agreement with experiment;
however they  have concluded that one electron models are inadequate to
explain infrared reflectivity data.  Nevertheless,    these results
have left open the possibility that the true ground state is
continuously connected to the one electron state as the
electron-electron interactions measured by a Hubbard $U$ are turned
on.  \cite{Doniach94}

In this paper, we construct a simple model Hamiltonian which appears
to contain the physical ingredients needed for understanding the data.
We examine our Hamiltonian, a two band Hubbard model, in the limit of
large dimensions.  This limit enables us to approximately calculate properties
which can be compared to angle resolved photoemission experiments    and
 magnetic
susceptibility measurements.  In spite of the fact that the large
$D$ limit is effectively a local one, our model predicts strong
momentum dependence of the quasiparticle lifetime which should be observable
in photoemission experiments, as well as the usual band renormalization
effects one expects to find as a result of the correlations.  Despite
the simplified nature of the model we are able to account for the
observed temperature dependence of the magnetic susceptibility and of
the optical reflectivity at a semi-quantitative level.

\section{A Model Hamiltonian}
\subsection{Rationale}
Several experimental results guided the structure of our model.  The
model density of states constructed by Jaccarino {\it et
al.}\cite{Jaccarino67} to fit their susceptibility and specific heat
data required a narrow bandwidth of order 50 meV or less for a
reasonable fit which is unphysically small without renormalization
effects.  Schlesinger {\it et al.}~observed a strong temperature
dependence of the gap in the optical measurements, which could be
ascribed to a loss of coherence.  Such effects occur at half filling in
mean field solutions of the Anderson lattice Hamiltonian, a focus of
studies on cerium based strongly correlated insulators (e.g. Ref.
\onlinecite{Doniach94} and references therein).
However, we do
not expect to find the extremely narrow bands characteristic of
$f$-states in a transition metal compound such as FeSi, and indeed, the
only anomaly in the specific heat of FeSi
 matches well with  FeSi's magnetic susceptibility and  lacks the heavy
fermion behavior associated with the Anderson lattice.

LDA calculations for FeSi show a
nearly direct insulating gap, surrounded by several narrow iron $d$ bands
with rather small contributions from the silicon orbitals.
Thus we were led to construct the following two band Hamiltonian:
\label{twohubb}
\begin{eqnarray}
H = \sum_{<ij> \sigma } -t ( \hat c^\dagger_{i 1 \sigma }  \hat c_{j 1 \sigma }
- \hat c^\dagger_{i 2 \sigma }  \hat c_{j 2 \sigma }) \cr
 + \sum_{i \sigma } v  (\hat c^\dagger_{i 1 \sigma } \hat c_{i 2 \sigma }
 + h.c.)
 + \sum_{i} U(\hat n_{c_{i 1\uparrow}} \hat n_{c_{i 1\downarrow}}
 + \hat n_{c_{i 2\uparrow}} \hat n_{c_{i 2\downarrow}}).
\end{eqnarray}
Here we have coupled two
Hubbard bands $\hat c_{i j \sigma }$, which may be thought of as the
iron $d$ bands,  with a momentum independent hybridization $v$.
Choosing hopping terms of opposite sign provides a simple direct gap.

In contrast to the Anderson lattice Hamiltonian's conduction band and
strongly correlated $f$-band, our model has two symmetric tight binding
bands, with a Hubbard $U$ term acting within each band independently.
Note that a model with an additional on site repulsion $U' /2
(\hat n_{c_{i 1}} + \hat n_{c_{i 2}} ) (\hat n_{c_{i 1}} + \hat n_{c_{i 2}} -1)
$ would shift the energy level at half filling to a middle Hubbard
band, leaving the intraband $U$ to dominate the physics.  For simplicity
we
do not include  $U'$ in the  model.  Instead, our form of intraband $U$
focuses on the   competition between $U$ and $v$.  For $U=0$ the two
bands mix  and create a direct gap of magnitude $2v$.  The ground state
simply consists of doubly occupying the states in the lower hybridized
band.  As $U \rightarrow  \infty $ the electrons half fill   each orbital
to avoid double occupancy.

\subsection{Further Approximations: $D=\infty$
\& Self Consistent Perturbation Theory}

The large $D$ approximation, where $D$ is the number of dimensions,
 introduced by Metzner and Vollhardt\cite{Metzner89} to study the
Hubbard model, naturally supports several common approximations, (e.g.
the ``local'' approximation , Gutzwiller's approximation, and some
slave boson results).  These approximations underlie our understanding
of the heavy fermions, and so we chose the large $D$ limit for study.
Using self consistent second order perturbation theory for small and
intermediate $U$,
Schweitzer and Czycholl\cite{Schweitzer90b}
 found only minor corrections in $1/D$ for the
Anderson lattice Hamiltonian for $D=2$, and found that the corrections
compared well with exact results for $D=1$. We chose to
follow their approach, as FeSi's structure is
three dimensional, and the renormalization effects we seek are  moderate.

The first major simplification comes in the unperturbed density of
states.  The energies of the two hybridized tight binding bands are
$\tilde\epsilon ({\bf k}) = \pm \sqrt {v^{2} + \epsilon ({\bf
k})^{2}}$, where $\epsilon ({\bf k}) = -t \sum_{i=1}^D \cos(k_i)$.  To
keep the  bandwidth of the unhybridized states the same, we should set
$t=t^*/\sqrt {2 D}$.  This is  easily seen\cite{MullHart89a} by
considering the density of states for a  single tight binding band,

\begin{eqnarray}
\rho _{TB}(\varepsilon) =  \prod_{i=1}^D \int _{-\pi }^\pi  {dk_i \over 2 \pi }
	\delta (\varepsilon - \sum_j^D -2 t \cos k_j)\cr
= \int {d\tau   \over 2\pi } e^{i \varepsilon \tau} \prod_{i=1}^D
	\int _{-\pi }^\pi  {dk_i \over 2 \pi } e^{-2i t \tau  \cos k_i}\cr
=\int  {d\tau   \over 2\pi } e^{i \varepsilon \tau} \exp(-2 D \tau ^{2} t^{2}
	+ O(t^4)) .
\end{eqnarray}
By performing a cumulant expansion in $\tau $, the density of states is
reduced to a Gaussian to leading order in $D$ for a single band.  We
measure all energies in units of  $t^*$, and then for $U=0$, we exactly
diagonalize the Hamiltonian.  The resulting density of states is
singular, with a gap of $\Delta _0=2v$, (in contrast to the indirect
gaps $ \delta \propto {v_R}^2/D$  in the Anderson lattice Hamiltonian
where $v^R$ is a renormalized hybridization and $D$ the bandwidth)

\begin{equation}
\rho _0(\varepsilon) = {1 \over \sqrt \pi } {\varepsilon \over
	\sqrt {\varepsilon^{2}-v^{2}}} \exp(-\sqrt {\varepsilon^{2}-v^{2}})
	(|\varepsilon| > v).
\label{eq:dos}
\end{equation}

For $U \not= 0$ we incorporate the effects of interactions on   the
density of states, as well as other properties, by calculating the
self-energy ${\bf \Sigma }({\bf k}, \omega )$ of the particles. This
determines the full Green's function via Dyson's equation,
\begin{equation}
\label{dyson}
{\bf G}({\bf k}, \omega ) = ({\bf G}_0^{-1}({\bf k}, \omega ) -
	{\bf \Sigma }({\bf k}, \omega ))^{-1}.
\end{equation}
Searching for a self consistent solution for this in perturbation
theory, we simplify the form of ${\bf \Sigma }({\bf k}, \omega )$ by
noting we can choose $\Sigma _{11} = \Sigma _{22}$ and $\Sigma _{12} =
\Sigma _{21}$, because $\Sigma _{ij}$ will, to second order, depend
only on $G_{ij}$, and $\epsilon _1({\bf k}) = -\epsilon _2({\bf k})$.

The first order terms in the expansion correspond to Hartree-Fock
theory; they simply represent the effect of a mean background field
upon the particles.  If we assume a paramagnetic state, then at half
filling $ \Sigma ^{(1)}_{ii,\sigma } ({\bf k}, \omega ) = U n_{i
\bar\sigma } = U/2 $.

For the second order terms we take advantage of the large $D$ limit, in which
the self energy becomes momentum independent.
A standard diagrammatic expansion yields,
\begin{eqnarray}
\label{sig2}
\Sigma ^{(2)}_{ij,\sigma } ({\bf k}, i\omega _n) = -U^{2}\beta ^{2}
\sum_{i\omega _1, i\omega _2,{\bf k}_1 \atop {\bf k}_2, {\bf k}_3}
G_{ij,\bar\sigma } ({\bf k}_2, i\omega _1) \cr
G_{ij,\bar\sigma } ({\bf k}_3, i\omega _2)\cr
 G_{ij,\sigma } ({\bf k}_1, i\omega _n -i\omega _1 + i \omega _2)\cr
\delta ({\bf k}_1 + {\bf k}_2 - {\bf k}_3 - {\bf k})\cr
=  -U^{2}\beta ^{2} \sum_{i\omega _1, i\omega _2} \int d\epsilon _1
	d\epsilon _2 d\epsilon _3 G_{ij,\bar\sigma } (\epsilon _2, i\omega_1)\cr
  G_{ij,\bar\sigma } (\epsilon _3, i\omega _2) G_{ij,\sigma }
  (\epsilon _1, i\omega _n -i\omega _1 + i \omega _2)\cr
w({\bf k}; \epsilon _1, \epsilon _2, \epsilon _3),\cr
w({\bf k}; \epsilon _1, \epsilon _2, \epsilon _3) =
	\sum_{{\bf k}_1 {\bf k}_2 {\bf k}_3}\delta ({\bf k}_1 + {\bf k}_2
	- {\bf k}_3 - {\bf k}) \cr
 \prod_{i=1}^3 \delta (\epsilon _i - \epsilon ({\bf k}_i)),
\end{eqnarray}
If we express the delta functions in  $w({\bf k}; \epsilon _1, \epsilon
_2, \epsilon _3)$ as integrals and sums over exponentials as before, we
find that to leading order in $1/D$, $w({\bf k}; \epsilon _1, \epsilon
_2, \epsilon _3)=1$, and thus $\Sigma _{ij}$ is momentum independent.
This result is true to all orders in perturbation
theory.\cite{MullHart89a}

In evaluating the second order corrections to $\Sigma _{11}(\omega )$
and $\Sigma _{12}(\omega )$ (again, assuming a paramagnetic state), the
Lehmann spectral representation of the Green's function,
\begin{eqnarray}
\label{lehmann}
G_{ij}({\bf k},z) = -{1 \over \pi}\int_{-\infty }^\infty
{ d\omega  \mathop{\rm Im}\nolimits G_{ij}({\bf k}, \omega  + i \epsilon )
\over {z - \omega }},
\end{eqnarray}
defines the Green's function over the complex plane in terms of its
values for real frequencies.  Our expression for $\Sigma ^{(2)}(\omega)$
only depends on the  on-site real space component, so we define,
\begin{eqnarray}
\bar G_{ij}(\omega  + i \epsilon ) = \sum_{{\bf k}}
	\mathop{\rm Im}\nolimits G_{ij}({\bf k},\omega  + i \epsilon ),\cr
= {1 \over \sqrt {\pi }} \int _{-\infty }^{\infty } d \epsilon _{{\bf k} }
	\exp(-\epsilon _{{\bf k}}^2) \mathop{\rm Im}\nolimits
	G_{ij}(\epsilon _{{\bf k}} ,\omega  + i \epsilon ).
\end{eqnarray}
$\bar G_{ij}(\omega  + i \epsilon ) $  can be evaluated  in terms
of Fadeeva's function (Appendix A), which has  generally available
numerical implementations.\cite{Poppe90}
Then we evaluate the sums over
Matsubara frequencies,
\begin{eqnarray}
\Sigma ^{(2)}_{ij} (i\omega _n) = -{U^{2} \over \pi ^3} \int   dz_1
dz_2  dz_3  \bar G_{ij}(z_1) \bar G_{ij}(z_2) \bar G_{ij}(z_3)\cr
\times { f(z_1) f(z_2) (1 - f(z_3)) + (1-f(z_1))(1-f(z_2))f(z_3) \over
z_1 + z_2-z_3- i\omega _n }.
\end{eqnarray}
If we analytically continue the self energy $(i \omega _n \rightarrow
\omega  + i\epsilon )$ then the energy denominator can be rewritten as
an integral over  an exponential.
 The resulting expression couples functions of the Green's functions
only through a Fourier transform which we evaluate numerically:
\begin{eqnarray}
\label{selfeq}
\Sigma ^{(2)}_{ij} (\omega ) = -{iU^{2} \over \pi ^{3}}
	\int _0^\infty  d\tau  e^{i \omega \tau }
	(\alpha _{ij}(\tau ) \alpha _{ij}(\tau ) \beta _{ij}(-\tau ) \cr
+ \beta _{ij}(\tau ) \beta _{ij}(\tau  )\alpha _{ij}(-\tau )).
\end{eqnarray}
Here $\alpha _{ij}(\tau )$, and~$\beta _{ij}(\tau )$ are Fourier
transforms of the site diagonal
Green's functions,
\begin{eqnarray}
\alpha _{ij}(\tau ) = {-1 \over \pi}\int _{-\infty }^{\infty }
	d\varepsilon~e^{-i \tau  \varepsilon}
	\bar G_{ij}(\varepsilon + i\epsilon ) f(\varepsilon),\cr
\beta _{ij}(\tau ) = {-1 \over \pi}\int _{-\infty }^{\infty }
	d\varepsilon~e^{-i \tau  \varepsilon}
	\bar G_{ij}(\varepsilon + i\epsilon )  (1-f(\varepsilon)).
\end{eqnarray}

To solve these equations self consistently, we start with a null self
energy, and calculate the corrections according to  eq. \ref{selfeq}.
The succeeding values of the self energy are linear combinations of the
current value and the new calculation.  By gradually mixing in the new
solution, we avoid oscillatory behavior.  When mixing in 20\% of the
new solution with the old, we find convergence within 10 iterations.
Our choice of mesh size for the discrete Fourier transform produced
results consistent with larger mesh sizes.
The evaluation of the Green's function at frequency
$ \omega  + i \epsilon $ in eq.\ref{dyson} is performed numerically; by keeping
$i\epsilon $ finite we force a finite lifetime for all states.

With a self consistent $\Sigma _{ij}(\omega )$ we  calculate several
physical quantities.  The  spectral function
$A(\epsilon _{{\bf k}}, \omega ) = -{1 \over \pi } \mathop{\rm Im}\nolimits
G(\epsilon _{{\bf k}}, \omega )$
and the one-particle density of states
$\rho (\omega ) = \sum_{\bf k} A(\epsilon _{{\bf k}}, \omega )$ can be compared
 to angle resolved and angle  integrated photemission studies.  The
static and dynamic susceptibilities are also easily calculated in the
large $D$ limit.  For all momenta except for ${\bf q}=0$ and ${\bf q}=
{\bf Q}$, if we exclude vertex corrections, the momenta decouple and
the two particle response function is simply the product of the one
particle functions.  In the case of the optical conductivity, this
decoupling, along with the parity of the current operator were shown by
Khurana\cite{Khurana90} to force the vertex corrections to vanish.
Thus to compare our model with optical conductivity measurements, we
need only compute the joint density of states:
\begin{equation}
\sigma (\omega , T) = \int d\epsilon  \rho (\epsilon )  \rho (\epsilon +\omega
) (1- f(\epsilon +\omega , T))
\end{equation}
While vertex corrections are nonzero for the magnetic
susceptibility\cite{Zlatic90}, for our system we expect no magnetic
instabilities, and so approximate
the magnetic susceptibility $\chi (T)$ for small to to intermediate $U$
  as simply
\begin{equation}
\chi (T) \propto \lim_{\omega \rightarrow 0} \sigma (\omega ,T)/T.
\end{equation}

\section{Results}
Our model  has two adjustable parameters,
$v$ and $U$ measured in units of the hopping parameter $t^*$. To
compare our model  with physical properties in FeSi, we use our LDA
calculation of the band gap ($\approx$ 150 meV) relative to the width
of the $d$ bands ($\approx$ 800 meV) to  choose $v=0.125$.  Our choice
of $U$ is based upon consistency between our results and experimental
observations.

Figure \ref{fig:self} depicts  a calculation of the self energy for a
range of temperatures.  Our calculation satisfies Luttinger's theorem,
as $\mathop{\rm Im}\nolimits \Sigma _{11}(\omega ) =0 $  at the Fermi
energy at zero temperature.  While calculating the corrections to the
self energy is the heart of our calculation, it is difficult to compare
to experimental results, so we consider quantities such as the DOS.

{}From  the density of states we see the gap is preserved,
but reduced by the interactions at zero temperature.
Because the boundary of the gap is  singular for $U=0$ (eq. \ref{eq:dos}),
we define the gap size as the separation between
the two maximal values in the DOS.  We find a fair empirical fit for
the gap $\Delta $,  as a function of $v$ and $U$,
\begin{equation}
\Delta (v,U) = \Delta _0  v^r/v,
\label{eq:fitgap}
\end{equation}
where
\begin{equation}
 v^r = {v\over \sqrt {1+0.4U^{2}/v}}
\end{equation}
is a renormalized coupling (Fig.\ \ref{fig:gap}).
This confirms our expectations of a competition between
$U^{2}$ (because corrections are second order in U)  and $v$ to reduce the gap.

If we look at the spectral function for a fixed value of $\epsilon
_{\bf k}$ close to the band edge,  we find a familiar scenario
(Fig.\ \ref{fig:angres}).  Our model exhibits a  quasiparticle pole,
with an incoherent background of width $\approx U$ forming above a
threshold at $v$.  By  numerically fitting the quasiparticle pole to a
Lorentzian, $ \alpha  \Gamma /((\omega -\omega _0)^{2} + \Gamma ^{2})
$, we can study the dependence of the quasiparticle's energy $\omega
_0$, scattering rate $\Gamma $, and weight $\alpha $ on $\epsilon _k$
and $T$.

Examining the dispersion of the quasiparticle energy, we find it is
strongly renormalized by $U$, and that the overall bandwidth is reduced
by roughly $v^r/v$   (Fig.\ \ref{fig:quasi-disp}).  For moderate
temperatures, the temperature dependence of this dispersion is
relatively minor (Fig.\ \ref{fig:disp-t}).

The scattering rate  of the quasiparticles near the Fermi energy
exhibits a  simple temperature dependence, with $\Gamma  \propto
\exp(-\beta \Delta )$.  Thus the lifetime of a quasiparticle is
determined by scattering off of thermally excited particles and holes
near the fermi energy.  However, this scattering rate is strongly
dependent on $\epsilon _k$.  As the quasiparticle energies reach the
energy scale of particle-hole pair creation, the lifetime dramatically
shortens (Fig.\ \ref{fig:gamma-t}).

The distribution of spectral weight between the quasiparticle peak and
the incoherent background is strongly temperature dependent
(Fig.\ \ref{fig:angres}).  At low temperatures, the weight is
 divided between the two and remains roughly constant. However, as the
temperature increases, the spectral weight rapidly shifts to the
broadened quasiparticle pole until that is all that remains visible.
While there is some uncertainty in the magnitude of this effect as a
result of our fitting procedures, it is strong enough that the shift in
spectral weight is clearly established.

%
The two particle Green's functions provide other tests of our model.
The static susceptibility, which we approximate by $\chi (T) = \sigma
(0,T)/T$ provides a nice illustration of the effects of the on site
correlation.  In Figure \ref{fig:susc} we have plotted $\chi (T)$ for
the interacting and non-interacting cases ($U=0$ and $U=1$)
respectively, and compared them to experimental data.\cite{Tajima88}
Both cases exhibit a gap which eventually turns over to a Curie form
$\chi  \propto 1/T$, but the finite $U$ case saturates more quickly and
provides a much better fit to the data.

In Figure \ref{fig:susc} by scaling $\chi (T)$ we connect the energy
scales of our theory to experiment.  For this scale, the natural
energy unit, (which comes from the hopping   term $t$) corresponds
to an energy of 0.7 eV.  This leads to a  renormalized gap size of
about $70$ meV, and a non-interacting one of about $140$ meV
for the above choices of $v$ and $U$.
These correspond well to
 the experimental findings and to the LDA
calculations of FeSi, respectively.

Another experimental test is the optical conductivity.  In Figure
\ref{fig:pjdos} we show the joint density of states as a function of
temperature.  At low temperatures we have a gap indicating the
separation between the two quasiparticle bands and a secondary band
created by transitions to states in the background.  As we increase the
temperature, the conductivity smears out, and the spectral weight
becomes distributed over  a much larger range in energies.  While there
is some shift of spectral weight to within the gap, the reduction of
the main peak far exceeds this as it is distributed across all
temperatures.  The gap fills in uniformly, with significant filling by
$T \approx \Delta /2$, in contrast to models where  there are
 free carriers  in a Drude peak.\cite{Fu94}

\section{Discussion}
In this paper we have begun exploring the possibilities of a simple
model Hamiltonian for modelling  the narrow gap semiconductor FeSi,
using the large $D$ limit to compare with specific experimental
results.   The large $D$ limit provides a simple method of
incorporating the effect of electron correlations.  After choosing
model parameters for the base Hamiltonian consistent with our earlier
electronic structure calculations in the LDA, we calculated one and two
particle properties in self-consistent second order  perturbation
theory which we believe will provide a reasonable approximation at small to
intermediate values of $U$.

The many body effects modify the Bloch states resulting  in a
quasiparticle spectrum with an incoherent background.  The effects of
the correlations on the  quasiparticle peaks leads to a reduction of
the gap which, for moderate values of $U$ may explain the discrepancies
seen   between band structure calculations and observed properties of
FeSi.

The renormalization of   the quasiparticle peaks also manifests itself
as an apparent narrowing of the bands themselves. The resulting
paramagnetic susceptibility explains the observations of $\chi (T)$
which had long ago been fit to  an extremely narrow two band model.
While one might question the need to consider correlation effects given
the relatively good fit for the $U=0$ case in Fig.\ \ref{fig:susc}, one
must remember that even in the non-interacting state our infinite
dimensional model has  very narrow bands in the DOS (eq.
\ref{eq:dos}).  Nevertheless, the correlation reduced bandwidths are
more rapidly saturated as the temperature increases, resulting in a
more favorable comparison with experiment.

Several features from our calculation should be very apparent in angle
resolved photoemission experiments.
The lifetime and spectral weights of the
narrow quasiparticle peaks are strongly temperature dependent as a result
of particle-hole pair creation.  The
incoherent portion of the spectral function is also temperature dependent.
This could be potentially significant in analyzing photoemission spectra,
as it may bias attempts to normalize spectral weights at different
temperatures.

However, our model does not reproduce the shifts of spectral weight
seen by Schlesinger {\it et al.}~ in the optical conductivity. Thus
either the $D\rightarrow \infty $ limit is inadequate to account for
the shifts or there is some question about these experimental
findings.

Nonetheless, the simplicity of this model, a strongly correlated,
direct gap insulator, should lend itself well to a variety of
analytical techniques.  Future exploration could be made using
Gutzwiller projection operators or calculating
 finite $D$ corrections.
Georges and Kotliar\cite{Georges92} have shown that in the large $D$
limit the Hubbard model can be transformed into an impurity problem
with a supplementary self-consistency equation, and which can be solved
exactly, even in the limit of large $U$ (where the form of self consistent
second order perturbation theory used here  fails).
This inspired a great deal of work \cite{Georges94} which should
be extensible to our model as well.  All of these offer
opportunities to gain  further insights in the physics of strongly
correlated insulators.

\acknowledgments
We thank Z.X. Shen and C. H. Park for sharing
experimental results and providing very valuable feedback.
We thank Hidei Fukuyama  for an interesting
discussion. We thank Antoine Georges for helpful comments on
an earlier version of this paper.
This work was supported in part by the U.S. Department of
Energy under contract W7405-ENG-36 and in part by NSF grant DMR9302882.
\appendix
\section*{Local Green's function}
For a given frequency, we can evaluate the local Green's function
numerically in terms of Fadeeva's function,
\begin{equation}
\label{eq:fadeeva}
w(z) = {i \over \pi } \int _{-\infty }^\infty  {e^{-t^{2}} dt\over z-t}
(\mathop{\rm Im}\nolimits z > 0),
\end{equation}
which has a widely available computer implementation. \cite{Poppe90}
We start with
the Dyson equation, and replace the momentum summation by
an integration over the density of states,
\begin{eqnarray}
G_0^{-1}(\omega ; k) = \omega  I + \epsilon _k \tau _3 - v \tau _2,\cr
\bar G(\omega ) = \mathop{\rm Im}\nolimits \sum_{{\bf k}} (
	G_0^{-1}(\omega ;{\bf k}) - \Sigma (\omega ))^{-1},\cr
\bar G(\omega ) = \mathop{\rm Im}\nolimits {1 \over \sqrt \pi }
	\int _{-\infty }^{\infty } d\epsilon _k  e^{-\epsilon _k^{2}}
	(G_0^{-1}(\omega ;\epsilon _k) - \Sigma (\omega ))^{-1}\cr.
\end{eqnarray}
Because the self energy is independent of the momenta, the integrals
can be expressed in terms of $w(z)$, yielding,
\begin{eqnarray}
\bar G_{11}(\omega ) = \mathop{\rm Im}\nolimits {2\over \sqrt \pi }
	(z-\Sigma _{11}(\omega )) \zeta (\omega ),\cr
\bar G_{12}(\omega ) = \mathop{\rm Im}\nolimits {2\over \sqrt \pi }
	(\Sigma _{12}(\omega )-v) \zeta (\omega ),\cr
\zeta (\omega ) = -i \pi  {w\bigl(
	\sqrt {(\omega -\Sigma _{11}(\omega ))^{2}
	- (v-\Sigma _{12}(\omega ))^{2}}\bigr)
	\over (\omega -\Sigma _{11}(\omega ))^{2}
	- (v-\Sigma _{12}(\omega ))^{2}},
\end{eqnarray}
where we have taken the root within the domain of eq.~\ref{eq:fadeeva}.
\bibliographystyle{prsty}

\begin{figure}
\epsfxsize=3.4in
\epsffile{fig1.ps}
\nobreak
\caption{
Diagonal (bottom) and off-diagonal (top) components of the imaginary part
of the self energy
$\Sigma _ij(\omega )$  as a function
of temperature.  Solid, dashed, and dotted lines correspond to
T=0.33, 0.10, and 0.033 respectively.
}
\label{fig:self}
\end{figure}

\begin{figure}
\vbox{
\epsfxsize=3.4in
\epsffile{fig2-nopr.ps}
\nobreak
\caption{
Size of the gap $\Delta $, in the one particle spectral function
$-{1\over\pi } \mathop{\rm Im}\nolimits G(\omega )$ as a function of
$U$ for selected values of  $v$. $\Delta $
is the distance between maximal values of the spectral function.
Solid lines show  the empirical formula in
eq. \protect\ref{eq:fitgap}.
}
\label{fig:gap}
}
\end{figure}

\begin{figure}
\vbox{
\epsfxsize=7.0in
\epsffile{fig3-nopr.ps}
\nobreak
\caption{
One particle spectral function for $\epsilon _{\bf k}=0$,
$U=1, v=0.125$ at selected
temperatures: $T=0.067 \Delta _0 $ (a), $T=0.4 \Delta _0$ (b),
and $T=1.3 \Delta _0$ (c).
Dotted lines indicate a least squares fit of the quasiparticle
peak to a Lorentzian broadened by a Fermi function and  residuals
from incoherent scattering.  Lineshapes have been convolved
with a Gaussian  to reflect instrumental broadening.
}
\label{fig:angres}
}
\end{figure}

\begin{figure}
\epsfxsize=3.4in
\epsffile{fig4.ps}
\nobreak
\caption{
Inverse lifetime of the quasiparticle for $U=1, v=0.125$,
as a function of $T$ at $\epsilon _k = 0$, (a), and $\epsilon _k$,  (b).
}
\label{fig:gamma-t}
\end{figure}

\begin{figure}
\epsfxsize=3.4in
\epsffile{fig5.ps}
\nobreak
\caption{
Dispersion of the fitted quasiparticle
peak for  $U=1$(solid) and $U=0$ (dashed).
Dotted line illustrates the dispersion which would occur
if the only effect of $U$ were to renormalize $v$.
}
\label{fig:quasi-disp}
\end{figure}

\begin{figure}
\epsfxsize=3.4in
\epsffile{fig6.ps}
\nobreak
\caption{The $k$-dependence of the fitted quasiparticle energy for $U=1$ for
various temperatures.
Above $T=0.05$  the peak in the spectral function fits poorly to a Lorentzian,
and is unsuitable for comparison.
}
\label{fig:disp-t}
\end{figure}

\begin{figure}
\epsfxsize=3.4in
\epsffile{fig7.ps}
\nobreak
\caption{
Magnetic susceptibility $\chi(T)$ of  our model for $U=0.0$ (crosses),
and $U=1.0$ (plus signs), as a function of temperature.
The solid line indicates experimental measurements on FeSi from Jaccarino
{\it et al.}.\protect\cite{Jaccarino67}
}
\label{fig:susc}
\end{figure}

\begin{figure}
\epsfxsize=3.4in
\epsffile{fig8-nopr.ps}
\nobreak
\caption{Joint density of states $\sigma (\omega ,T)$.
Different curves indicate
temperature, for $v=0.125$ and $U=0.625$.
}
\label{fig:pjdos}
\end{figure}
\end{document}